\documentclass[pra,twocolumn,longbibliography,longbibliography]{revtex4-2}
\usepackage[T1]{fontenc}
\usepackage{graphicx}
\usepackage{amsfonts}
\usepackage{physics}
\usepackage{comment}
\usepackage{hyperref}
\usepackage{xcolor}
\usepackage{dsfont}

\newcommand{\mm}{\alpha}

\begin{document}
    \title{Bound entanglement in symmetric random induced states}
	\author{J.\ Louvet}
	\email{jlouvet@uliege.be}
        \author{F.\ Damanet}
        \email{fdamanet@uliege.be}
	\author{T.\ Bastin}
	\email{T.Bastin@uliege.be}
	
	\affiliation{Institut de Physique Nucléaire, Atomique et de Spectroscopie, CESAM, University of Liège, B-4000 Liège, Belgium}
	
	\begin{abstract}

    Bound entanglement, a weak -- yet resourceful -- form of quantum entanglement, remains notoriously hard to detect and construct. We address this in this paper by leveraging symmetric random induced states, where positive partial transpose (PPT) bound entanglement arises naturally under partial tracing when proper parameters are selected. We investigate the probability of finding PPT bound entanglement in symmetric random induced states constructed via two methods: partial tracing of symmetric multiqubit pure states on the one hand (MI) and tracing out a qudit ancilla on the other hand (MII). For $N > 3$ qubits, we demonstrate that bound entanglement naturally emerges under optimal parameters, with a probability of occurrence very close to 1. We show that the two methods produce different varieties of PPT bound entangled states, and identify the contexts in which each method offers distinct advantages. These methods provide a versatile toolkit for the generation of large families of random PPT bound entangled states without complex numerical optimization.
 
\end{abstract}

\maketitle

\section{Introduction} \label{sec:intro}
Since its discovery, quantum entanglement has been regarded as the most non-classical characteristic of quantum mechanics. Over the decades, extensive research efforts have been dedicated to the detection, characterization, and manipulation of quantum entanglement, which constitutes a pivotal resource for quantum sciences and technologies~\cite{HORODECKI2009, GUHNE20091}. Despite these efforts, determining whether a multipartite state is separable or entangled, a challenge known as the separability problem \cite{HORODECKI2009}, remains daunting in the most general scenario. For qubit-qubit and qubit-qutrit systems, there exists a straightforward necessary and sufficient condition to address this problem \cite{PERES1996, HORODECKI1996}. However, for systems in higher dimensions, the development of efficient analytical or numerical methods to solve the separability problem remain challenging. This difficulty is partially attributed to a form of entanglement that poses significant detection challenges, known as bound entanglement, which gathers all those states that are entangled but are not distillable, i.e., from which no pure maximally entangled pairs can be extracted using local operations and classical communication~\cite{HORODECKI1998}. The utility of these states has been demonstrated across various applications, including the conversion of pure entangled states \cite{ISHIZAKA2004}, quantum key distribution \cite{HORODECKI2008}, and quantum metrology \cite{PhysRevLett.120.020506,KAROLY2021}. Nonetheless, constructing such states remains a nontrivial task, given the incomplete characterization of the set of bound entangled states. 
The first bound entangled states were presented in \cite{HORODECKI1997,HORODECKI1998}, along with a family of states for which bound entanglement can be activated \cite{HORODECKI1999}. Subsequent analytical construction of bound entangled states has primarily employed unextendible product bases \cite{BENNETT1999, HALDER2019,Skowronek_2011,DiVincenzo2003}, and constructions of bound entangled states from existing ones have also been explored \cite{BEJ2021,BANDYOPADHYAY2005}. Prior work shows that bound entangled states can also be generated through Lorentz boosts \cite{CABAN023}. Numerous other analytical and numerical approaches have been investigated to construct bound entangled states \cite{AUGUSIAK2010,HALDER20192,WEI2010,RUTKOWSKI2019,PIANI2007,ZHAO2015,BADZIA2014,TURA2012,POPP2022,LI2021,SINDICI2018,CLARISSE2006603,MOERLAND2024}, but no straightforward efficient way to generate large families of bound entangled states has been investigated.


In this paper, we show how one can get bound entangled states within symmetric random induced states (RIS). A RIS is obtained by tracing out an ancillary system from an initial random pure state of the global system. Recent findings indicate that bipartite bound entanglement can be generically obtained in RIS for systems of sufficiently large dimensions ~\cite{AUBRUN2013}. However, the question remains to be studied and clarified for low dimensions. The present study addresses this gap by examining the efficiency of obtaining bound entanglement in multipartite symmetric $N$-qubit systems in RIS for low to moderate $N$ values. We study two distinct constructions that successfully generate bound entanglement with different properties. We show that both methods provide straightforward access to bipartite and symmetric multipartite bound entangled states, in contrast to more elaborated numerical techniques, since the only operation involved in generating random induced states is the partial tracing over an ancillary system.  

The paper is organized as follows. In Sec.~II, we first present the necessary background and definitions related to entanglement, bound entanglement, symmetric states, and RIS. In Sec.~III, we describe the two methods used for constructing bound entangled RIS states. In Sec.~IV, we present the results for both methods and compare them. Finally, in Sec.~V, we conclude and provide several perspectives on our work.

\section{Background} \label{sec:background} 
\subsection{Bound entanglement}

\begin{figure}
    \centering
    \includegraphics[width=0.8\linewidth]{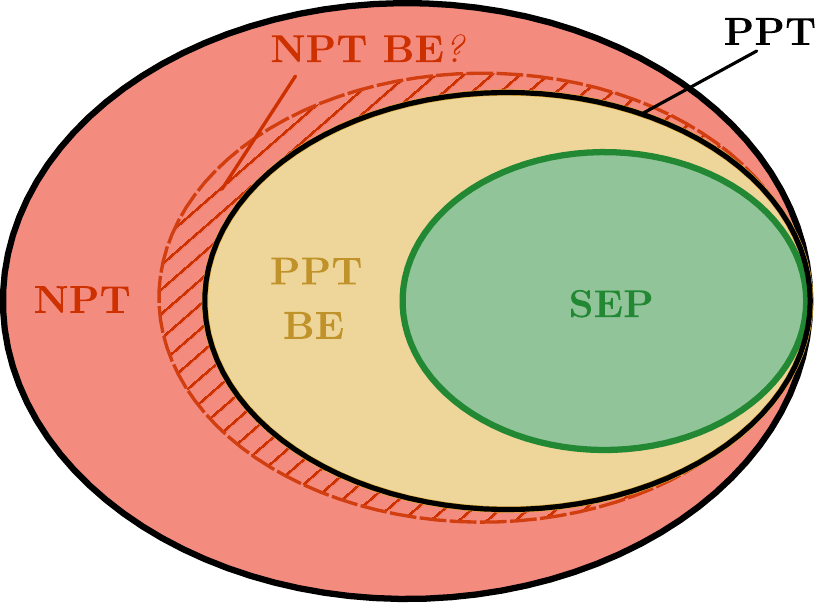}
    \caption{\textbf{Set of bipartite states :} Separable states (SEP), PPT bound entangled states (PPT BE), and NPT entangled states (NPT). The union of the PPT BE and SEP states yield all PPT states. It is still an open question whether there exists NPT bound entangled states (NPT BE) \cite{IQOQI}. For $2 \times 2$ and $2 \times 3$ systems, the PPT BE set is empty~\cite{PERES1996, HORODECKI1996}.}
    \label{fig:Set}
\end{figure}

\begin{figure*}[]
    \centering
    \includegraphics[width=0.7\linewidth]{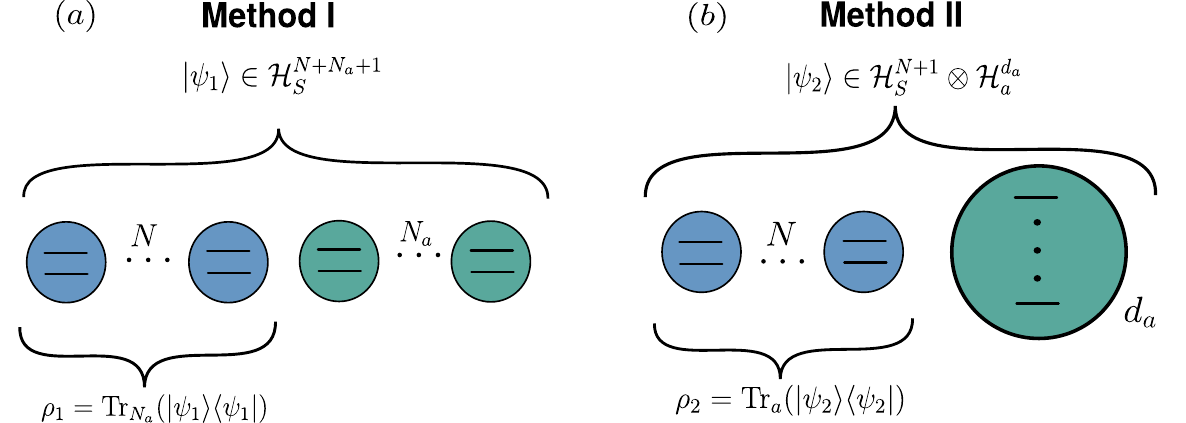}
    \caption{\textbf{Generation of RIS.} Method I (a): a RIS $\rho_1$ is created from an initial pure state $\ket{\psi_1}$ randomly generated on the symmetric subspace $\mathcal{H}_S^{N + N_a + 1}$ of $N+N_a$ qubits after tracing out $N_a$ qubits. Method II (b): a RIS $\rho_2$ is created from an initial pure state $\ket{\psi_2}$ randomly generated on the tensor product space $\mathcal{H}_S^{N+1} \otimes \mathcal{H}_a^{d_a}$ of the symmetric subspace of $N$ qubits and a $d_a$-dimensional qudit after tracing out the ancilla qudit.}
    \label{fig:RIS}
\end{figure*}

Entangled states $\rho$ are states that cannot be expressed as a convex combination of separable product states \cite{HORODECKI2009, GUHNE20091}. The Peres-Horodecki positive partial transpose criterion (PPT criterion) provides a straightforward sufficient condition for entanglement: a quantum state $\rho$ is entangled across the bipartition $A|B$ if its partial transposition $\rho^{T_A}$ has at least one negative eigenvalue. This criterion is necessary and sufficient for qubit-qubit and qubit-qutrit systems~\cite{PERES1996, HORODECKI1996}. In higher dimensions, PPT states, i.e., states with a positive partial transpose, may still exhibit entanglement. PPT entangled states have the property that they cannot be distilled \cite{HORODECKI19982}. The entanglement in non-distillable states is \textit{bound} to the state, hence their name: Bound Entangled (BE) states. By contrast, the question whether the entanglement of all negative partial transpose (NPT) states can be distilled remains open \cite{IQOQI}. Figure \ref{fig:Set} summarizes the different entanglement properties of bipartite states. Throughout this work, bound entangled states will refer to PPT BE states. In the context of multipartite systems, quantum states can be PPT w.r.t. some bipartitions being NPT w.r.t.\ others. To fully characterize a multipartite state, one thus has to go over every possible bipartition of the system. 

\subsection{Symmetric $N$-qubit states}
Multipartite symmetric states are states that remain invariant under any permutation of their parties. Symmetric $N$-qubit states live in the symmetric subspace $\mathcal{H}_S$ of dimension $N+1$. It is spanned by the Dicke states
\begin{equation}
\label{Eq.Dicke}
    \ket{D^{\mm }_N} = \binom{N}{\mm}^{-1/2} \sum_{\sigma} P_{\sigma}\big| \underbrace{0\dots 0 }_{N-\mm}   \underbrace{1 \dots 1}_{\mm} \big>
\end{equation}
for $\alpha = 0, \dots , N$, where the sum runs over all possible permutation operators $P_{\sigma}$ of the $N$ qubits. Two major simplifications emerge when investigating the entanglement of symmetric states. First, a symmetric state is either fully separable or genuinely multipartite entangled. Secondly, for any bipartition $A|B$ of the $N$-qubits, the explicit tracking of the qubits becomes redundant since any permutation of the qubits leaves the state unchanged. Only the number of qubits $k|N-k$ in the respective partitions $A|B$ is relevant. Hereafter, the bipartitions of symmetric qubits is denoted by their cardinality $|A|||B| \equiv k|N-k$.

\subsection{Random induced states}

A random induced state (RIS) on $\mathcal{H}^d$, where $d$ denotes the dimension of the state space $\mathcal{H}$, is defined as a mixed state obtained after adding an ancilla to the system and partial tracing a random pure state on the enlarged space $\mathcal{H}^d \otimes \mathcal{H}_a^{d_a}$ over the ancilla degrees of freedom, where $\mathcal{H}_a^{d_a}$ is the ancilla state space of dimension $d_a$. If the system state space $\mathcal{H}^d$  can be evenly bipartitioned as $\mathcal{H}^{d/2} \otimes \mathcal{H}^{d/2}$, it has been demonstrated that, for large dimensions $d$, the probability of getting an entangled RIS is high, as long as the ancilla state space dimension $d_a$ is smaller than some threshold $s$~\cite{AUBRUN2013}. Similarly, the probability of getting a PPT RIS is high, as long as $ d_a$ is greater than some other threshold $p < s$ \cite{AUBRUN2012}. Inbetween these two thresholds (for $p < d_a < s $), there is a high probability that the RIS is PPT entangled. While the arguments presented are also valid for unbalanced bipartitions, it is unknown if there exists such thresholds for low dimensions $d$, nor if a region of PPT entangled exists. We study here the entanglement of RIS of symmetric $N$-qubit states for all bipartitions, which has not been explored yet.

\section{Methods}

In this section, we present the two methods investigated to generate symmetric random induced states, as well as the method we employ to detect bound entanglement, and how we estimate the probability to get bound entanglement in the RIS. 

\subsection{Generation of RIS} 

We consider two different methods to generate random induced states:
\begin{itemize}
    \item Method I (MI): we generate a random $(N + N_a)$-qubit symmetric pure state $\ket{\psi_1}$ on $ \mathcal{H}_S^{N + N_a + 1}$ and trace out $N_a$ qubits, i.e., the RIS is given by $\rho_1 = \text{Tr}_{N_a} (\ketbra{\psi_1})$.
    
     \item Method II (MII): we generate a random pure state $\ket{\psi_2}$ on $\mathcal{H}_S^{N+1} \otimes \mathcal{H}_a^{d_a}$ and trace out the qudit ancilla system $\mathcal{H}_a^{d_a}$ of dimension $d_a$, i.e., the RIS is given by $ \rho_2 = \text{Tr}_{a} (\ketbra{\psi_2})$.
\end{itemize}

In both methods, the generated RIS $\rho$ are symmetric, i.e., $P_\sigma \rho = \rho = \rho P_\sigma$, for all permutations $\sigma$ of the $N$ qubits. For both cases, the initial random pure states $\ket{\psi_1}$ and $\ket{\psi_2}$ are generated according to the unitarily invariant Fubini–Study measure in the space of pure states following Ref.~\cite{ZYCZKOWSKI2011}. 
Figure~\ref{fig:RIS} illustrates both methods.

\subsection{Detection of PPT bound entanglement} 

In order to assess whether the generated RIS is NPT, PPT BE or SEP, we check if the state is PPT across the different bipartition of the $N$ qubits. Since the RIS is symmetric, as soon as one of its bipartition is NPT, the state is necessarily genuinely entangled, i.e., entangled accross every bipartition. If the state is in contrast PPT accross every bipartition, the separability problem must be solved in order to assess or not the full separability of the state. To do so, we use the reformulation of the separability problem as a truncated moment problem that we solve using semi-definite optimization~\cite{BOHNET2017}, as it is very efficient for symmetric qubits.  The different outcomes are then the following

\begin{itemize}
	\item \textbf{NPT} : The state is NPT for every bipartition. There is NPT entanglement across all partitions.
	\item \textbf{PPT BE} : The state is PPT for at least one bipartition and NPT for the others. There is PPT bound entanglement across the PPT bipartitions.
	\item \textbf{SEP} : The state is PPT for every bipartition and is fully separable. 
\end{itemize}

It may happen that in some cases the separability problem is unresolved with this method. These cases are tracked in our process and tagged UNK for unknown states. Note that for $N =2$ and $N =3$, the only bipartitions are $1|1$ and $2|1$ respectively. For these dimensions PPT states are also separable by the PPT criterion. PPT bound entangled bipartitions can then arise for $N>3$.

\subsection{Estimation of the probabilities} 

The probability $P_\mathrm{K}$ (K = NPT, PPT BE, SEP) of getting NPT, PPT BE, and SEP states, respectively, using either method I or II is given by 
\begin{equation}
    P_\mathrm{K} = \lim_{n\rightarrow\infty} \frac{n_\mathrm{K}}{n},
\end{equation}
with $n_\mathrm{K}$ the number of times a K state is found among the set of $n$ RIS using the respective methods. We define similarly the probability $P_\mathrm{UNK}$ that methods MI and MII generate a PPT RIS that cannot be detected either separable or entangled by our implementation of the truncated moment reformulation of the separability problem.

The probabilities $P_K$ were estimated using the empirical probabilities $\tilde{P}_\mathrm{K}(n) = n_K/n$ for a fixed number $n$ of RIS. For large $n$, $P_K \simeq \tilde{P}_\mathrm{K}(n)$. The validity of this approximation has been analysed through the convergence of $\tilde{P}_\mathrm{K}(n)$ with respect to $n$. We concluded that with $n = 10^5$ the approximation is expected to hold within a precision of $10^{-3}$ (see Appendix~\ref{app:conv}).  

\begin{figure*}
    \centering
    \includegraphics[width=1\linewidth]{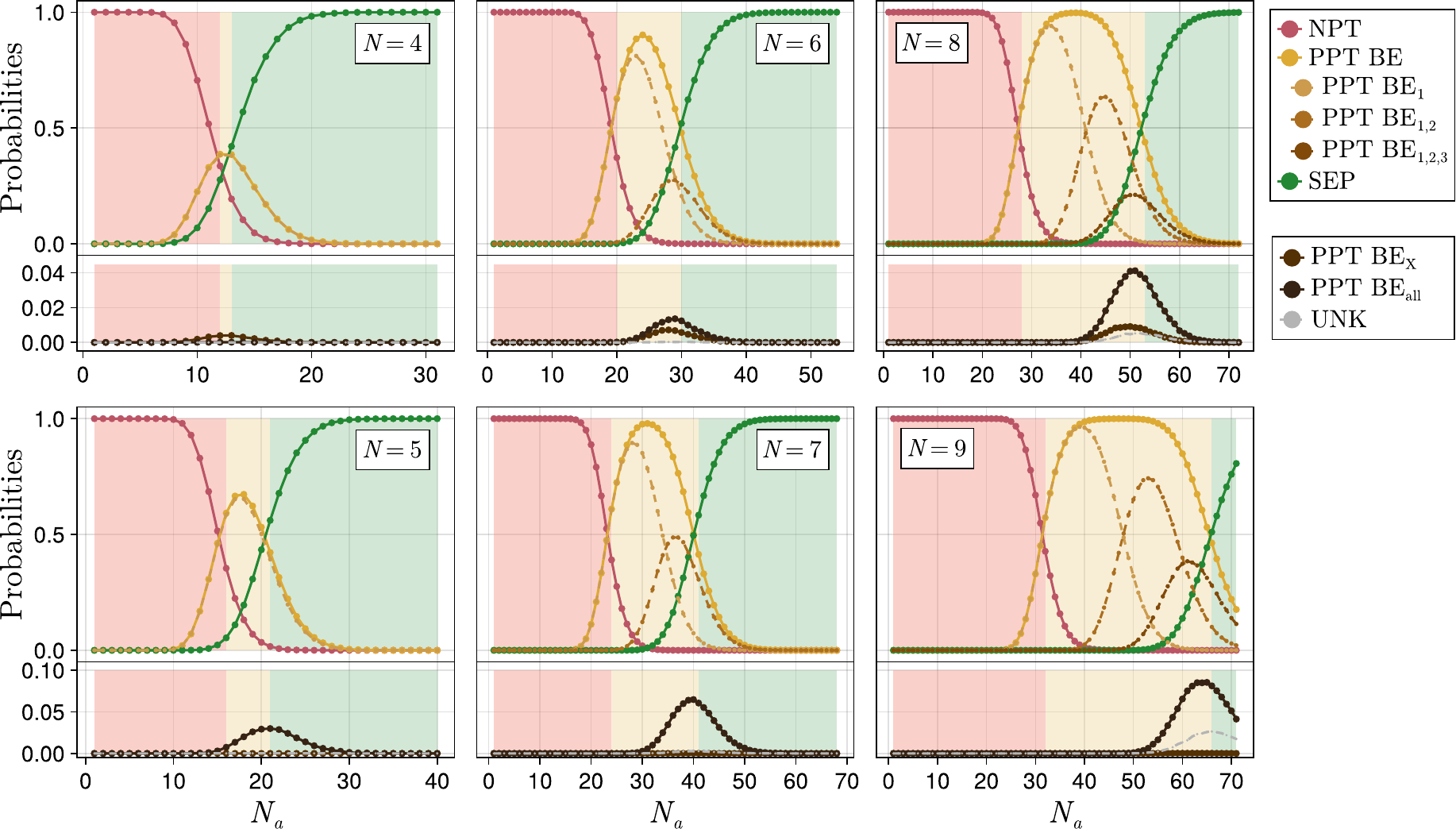}
    \caption{\textbf{Bound entanglement in Method I.} Probabilities that the generated random induced states have NPT entanglement (red), PPT bound entanglement (yellow), or are separable (green) using Method I with respect to the number of qubits $N_a$ in the ancilla system, for $N = 4,\dots,9$. $N_a$ starts at 1 and increases with a step of 1. The maximal PPT BE probability quickly grows to 1 as $N$ increases, showing the efficiency of the method. The brown curves are a refinement of the PPT BE yellow curve. They show the probabilities to get PPT bound entanglement across the different bipartitions: PPT BE$_1$, PPT BE$_{1,2}$, and PPT BE$_{1,2,3}$ as the ancilla parameter increases. The PPT BE$_1$ curve and the global PPT BE curve (yellow) are almost identical for $N=4$ and $5$. The bottom frame exhibit the 3 probabilities of getting PPT BE$_X$, PPT BE$_\mathrm{all}$, and UNK states. The lines between the dotted points are to guide the eye. The background colors indicate the regions where the probability to get NPT entanglement (red), PPT bound entanglement (yellow), or separable states (green) is the highest compared to the other two.}
    \label{fig:MethodI}
\end{figure*}

\begin{figure}
    \centering
    \includegraphics[width=1\linewidth]{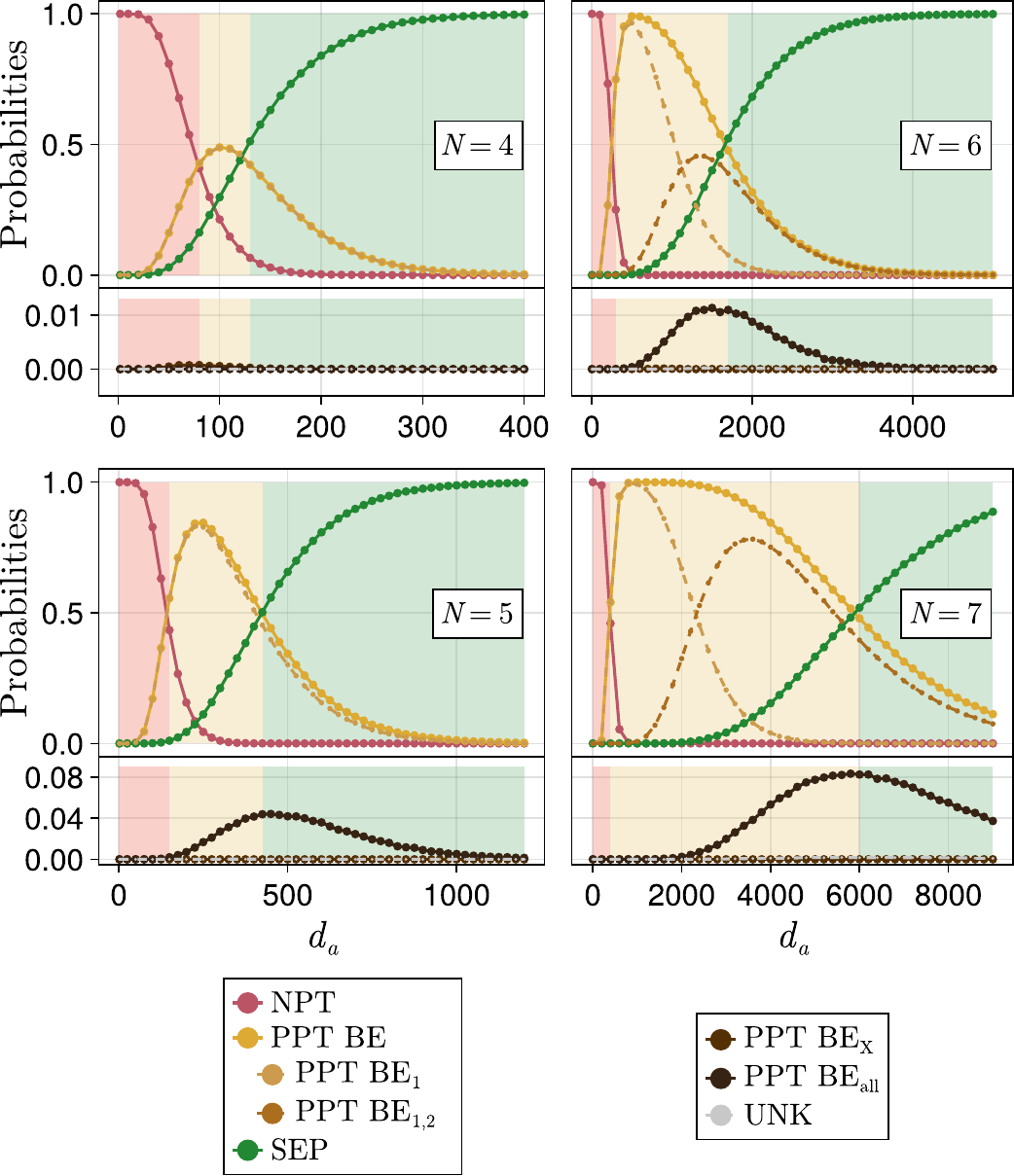}
    \caption{\textbf{Bound entanglement in Method II.} Probabilities that the generated random induced states have NPT entanglement (red), PPT bound entanglement (yellow), or are separable (green) using Method II with respect to the dimension of the ancilla qudit $d_a$ for $N = 4,\dots,7$. The value of the ancilla starts at $d_a = 2$ for each $N$. For $N = 4,\dots,7$, the following values of $N_a$ are multiples of $10, 25, 100$, and $200$, respectively. The brown curves are a refinement of the PPT BE yellow curve. They show the probabilities to get PPT bound entanglement across the different bipartitions: PPT BE$_1$ and PPT BE$_{1,2}$ as the ancilla parameter increases. The PPT BE$_1$ curve and the global PPT BE curve (yellow) are almost identical for $N=4$ and $5$. The bottom frame exhibit the 3 probabilities of getting PPT BE$_X$, PPT BE$_\mathrm{all}$, and UNK states. The lines between the dotted points are to guide the eye. The background colors indicate the regions where the probability to get NPT entanglement (red), PPT bound entanglement (yellow), or separable states (green) is the highest compared to the other two.}
    \label{fig:MethodII}
\end{figure}

\begin{figure}
    \centering
    \includegraphics[width=0.9\linewidth]{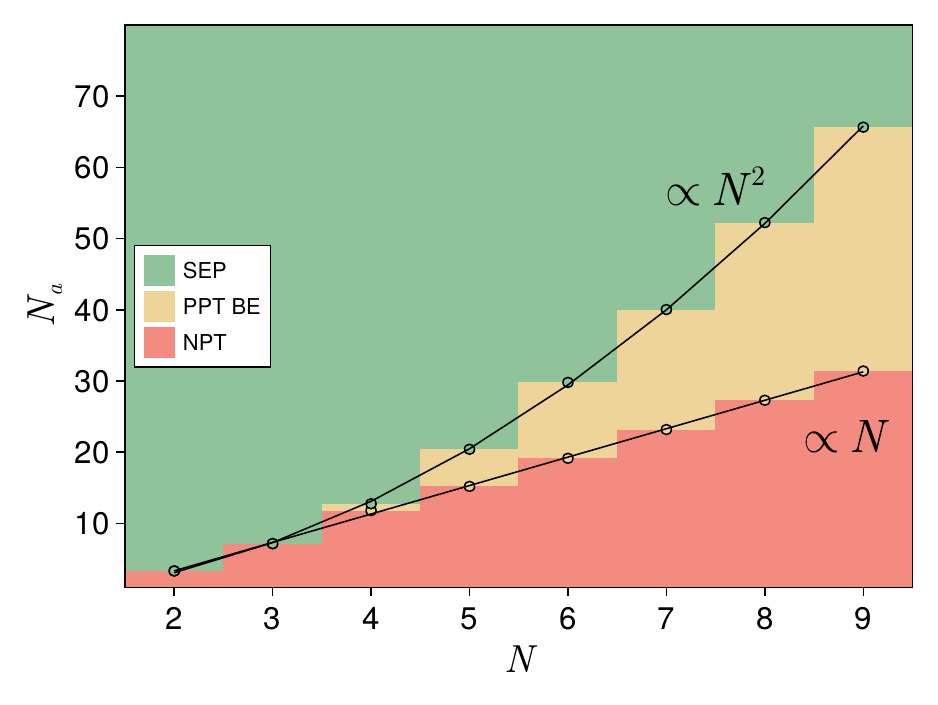}
    \caption{\textbf{Entanglement Phase Diagram.} Diagram highlighting where NPT (red), PPT BE (orange) and SEP (green) RIS are the most likely obtained for Method I, as a function of $N$ and $N_a$. The boundaries correspond to the intersections between the probability curves of figure \ref{fig:MethodI}. The lines are linear and quadratic fits of the intersection points. }
    \label{fig:PhaseDiag}
\end{figure}

\section{Results} \label{Results}

This section presents our results about the probabilities of generating PPT BE states, as well as NPT and SEP states, from the two kinds of RIS generated in this paper. We show how they scale with system size $N$. We also provide a discussion about the different properties of the obtained PPT BE states depending on the methods used to generate the RIS. 

\subsection{Probabilities}

As shown in Figs.~\ref{fig:MethodI} and \ref{fig:MethodII}, it is possible to create bound entanglement in symmetric $N$-qubit RIS with both methods MI and MII as soon as $N>3$. Upon increasing the ancilla parameters $N_a$ and $d_a$, we observe 3 distinct entanglement ``phases''. First, the probability to get a NPT RIS is very high (1), then decreases to reach 0, while the probability of getting a PPT BE state increases to reach a maximum. This maximum varies and increases with $N$ in MI for $N=4,5,6$ and in MII for $N=4,5$, reaching a value close to 1 in both methods for higher $N$. This probability then decreases to reach 0, while the probability to get a full separable RIS increases and becomes very high (1). This behavior aligns with the analytical results reported in~\cite{AUBRUN2012,AUBRUN2013} for large dimensions in the bipartite scenario. However here we do not observe sharp thresholds between the different phases. The increase (or decrease) in each probability is not abrupt, but rather gradual.

Let us focus on the PPT BE RIS (the yellow curves on Figure ~\ref{fig:MethodI} and ~\ref{fig:MethodII}). In order to ascertain precisely which bipartitions of the RIS exhibit PPT bound entanglement, it is necessary to consider every bipartition and every possible combination of them. The notation $\text{PPT BE}_{i_1,\dots,i_k}$ ($k \geq 1$) denotes that there is PPT bound entanglement across the bipartitions $\bigcup_{j=1}^{k} N-i_j|i_j$. In case the state is PPT bound entangled across all possible bipartitions [all possible bipartitions but the penultimate $\lceil N/2 \rceil + 1 | \lfloor N/2 \rfloor - 1$], it is also denoted by PPT BE$_\mathrm{all}$ [PPT BE$_X$]. 

For instance, for $N=6$, the different possibilities are $\text{PPT BE}_1$ (PPT BE across the bipartition 5|1), $\text{PPT BE}_2$ (across the bipartition 4|2), $\text{PPT BE}_3$ (across the bipartition 4|2) and 4 combinations of them: $\text{PPT BE}_{1,2}$ (across the bipartitions 5|1 and 4|2), $\text{PPT BE}_{1,3} = \text{PPT BE}_X$ (across the bipartitions 5|1 and 3|3), $\text{PPT BE}_{2,3}$ (across the bipartitions 4|2 and 3|3), and $\text{PPT BE}_{1,2,3} = \text{PPT BE}_\mathrm{all}$ (across the bipartitions 5|1, 4|2, and 3|3). 

We observed that out of all possible bipartitions, there is a non-zero probability of getting a RIS with PPT entanglement across only a few. In our example of 6 qubits, out of the 7 possible outcomes, there is a non-zero probability of getting a RIS with PPT entanglement across only 4 of them, and as the ancilla parameter increases, their respective probabilities start to increase in a specific order : $\text{PPT BE}_1$, $ \text{PPT BE}_{1,2}$, $\text{PPT BE}_{1,2,3}$ and finally $\text{PPT BE}_{1,3}$. These refinements can be observed on Figure ~\ref{fig:MethodI} and ~\ref{fig:MethodII}.

Generally, for an $N$-qubit system, as the ancilla parameters grow, PPT bound entanglement emerges progressively across more and more bipartitions. First appear $\text{PPT BE}_1$ states, then $\text{PPT BE}_{1,2}$ states, $\text{PPT BE}_{1,2,3}$ states and so on, until $\text{PPT BE}_\mathrm{all}$ states, and finally $\text{PPT BE}_X$ states. It occurs with a lower probability than $\text{PPT BE}_\mathrm{all}$, and does not behave similarly for every $N$, nor for the two methods. Indeed, $N=4$ is the only case where $\text{PPT BE}_X$ RIS are more likely to appear than $\text{PPT BE}_\mathrm{all}$ RIS in both methods. Moreover, the probability of $\text{PPT BE}_X$ is dramatically smaller (close to 0) for odd numbers than for even numbers $N$. In Method II., the probability of $\text{PPT BE}_X$ is 0 for odd numbers in our estimations. Another difference between odd and even $N$ can be observed in the growth of the maximal probability of  $\text{PPT BE}_\mathrm{all}$  with respect to $N$ as depicted in both figures ~\ref{fig:MethodI} and ~\ref{fig:MethodII}. Indeed, in both methods, the maximal probability increases from $N=4$ to $N=5$, but then decreases from $N=5$ to $N=6$, and increases again from $N=6$ to $N=7$, and so on, displaying a sawtooth behavior. Different behaviors between odd and even number qubits in symmetric qubits systems has also been observed in other works \cite{louvet24}.

\begin{figure}
    \centering
    \includegraphics[width=1\linewidth]{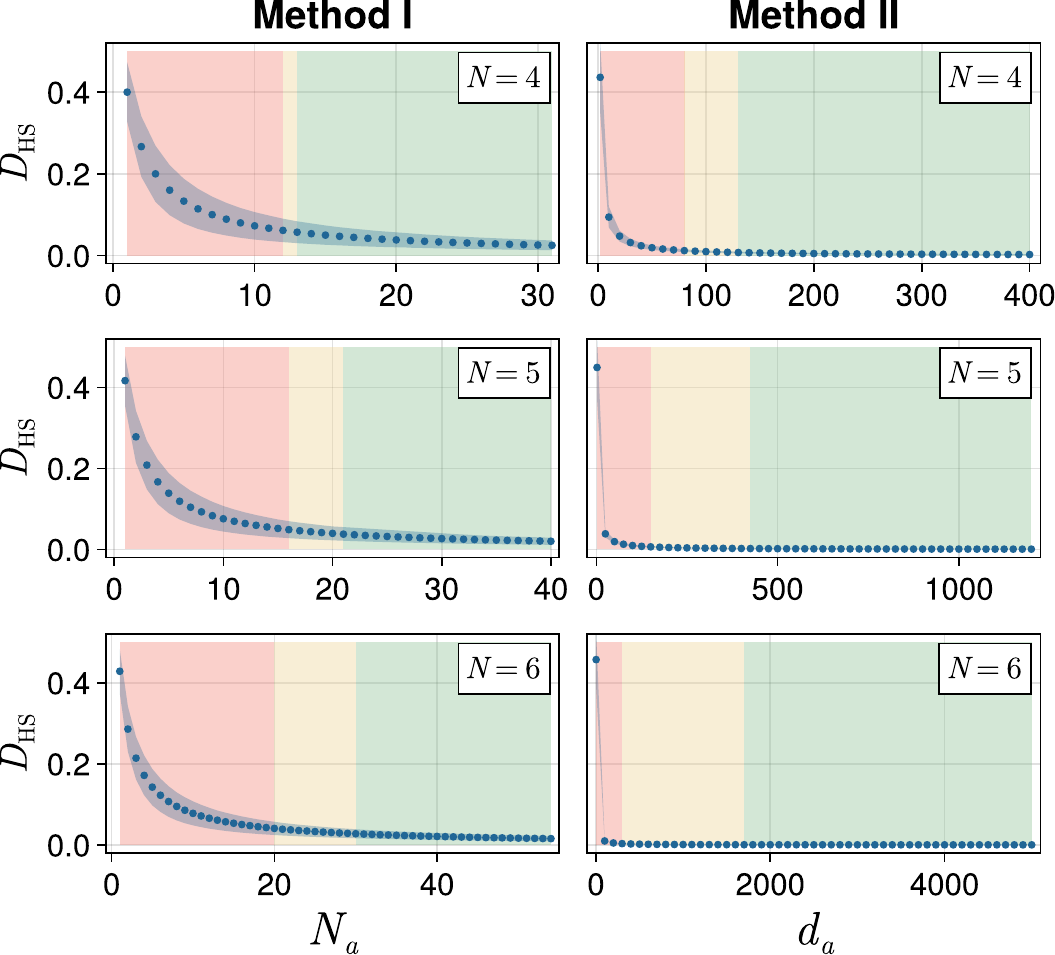}
    \caption{\textbf{Mean HS distance in RIS.} Mean Hilbert-Schmidt distance $D_\mathrm{HS}$  between the RIS and the maximally mixed state for $N=4,5,6$ as a function of the ancilla parameter ($N_a$ or $d_a$) for both methods MI and MII. For each parameter of the ancilla, 100 000 RIS were generated. The lighter blue band indicates the standard deviation, and is continuous to guide the eye. The background colors indicates the regions where the probability to get NPT entanglement (red), PPT bound entanglement (yellow), or separable states (green) is the highest compared to the other two.}
    \label{fig:purity}
\end{figure}

\begin{figure}
    \centering
    \includegraphics[width=1\linewidth]{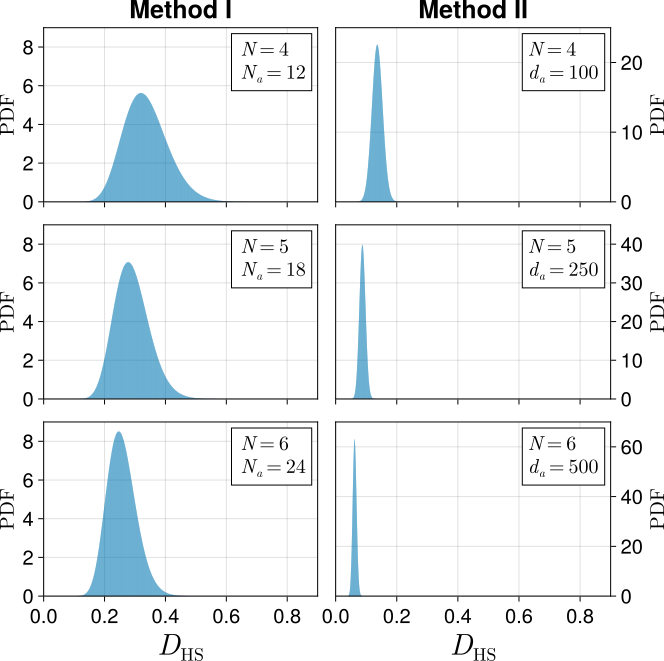}
    \caption{\textbf{Variety of bound entangled RIS.} Probability density functions (PDFs) of the Hilbert-Schmidt distance between two random induced bound entangled states using both methods MI and MII (and considering a set of 3 distinct parameters for each method - see insets). The comparison of the graphs shows that Method I presents a larger variety of bound entangled states than Method II.
    }
    \label{fig:DistBE}
\end{figure}

The range of the ancilla parameters $d_a$ and $N_a$ for which there is the highest probability to get PPT bound entanglement increases with the number of qubits $N$. By identifying the intersection points of the probability curves for distinct entanglement ``phase'' (NPT, PPT BE, and SEP) as a function of the ancilla parameter $N_a$ for each system size $N$, one can map out the dominant regimes of each phase. These intersections, extracted from Figure~\ref{fig:MethodI}, define boundaries in the resulting phase diagram (Figure~\ref{fig:PhaseDiag}). Notably, the transition from NPT to PPT exhibits a linear dependence on $ \sim N$, whereas the transition from entanglement to full separability follows a quadratic scaling  $\sim N^2$. This scaling behavior highlights that the regime in which bound entanglement can occur grows with $N^2$, underscoring its increasing relevance in larger systems.

\subsection{Hilbert-Schmidt distance and variety of PPT BE RIS}

 With the help of the Hilbert-Schmidt (HS) distance, we investigate here the variety of PPT BE RIS generated using the two methods. The Hilbert-Schmidt distance is defined between two operators $A$ and $B$ as $D_{\mathrm{HS}}(A,B) = \sqrt{\text{Tr} [ ( A - B)^{\dagger} ( A - B)] }.$ We first determined the mean distance between random induced states and the maximally mixed state (MMS) $\rho_0$ in the symmetric subspace, 
\begin{equation}
\rho_0=\frac{1}{N+1}\sum_{m=0}^{N} \ket{D^{(\mm )}_N}\bra{D^{(\mm )}_N}=\frac{\mathds{1}_{N+1}}{N+1},
\end{equation}
with purity $\trace(\rho_0^2) =  1/(N+1)$.  
Figure~\ref{fig:purity} depicts how the mean distance for MII decreases dramatically faster to 0 than for MI, for $N = 4,5,6$. The yellow background on the panels indicates where the probability to get PPT BE states is the highest compared to the probability to get NPT or SEP RIS. These observations indicates that MII creates bound entangled states much closer to the MMS compared to MI. The same behavior is observed for other numbers $N$ in both methods. Note that $D_{\mathrm{HS}}(\rho_S,\rho_0)$ is directly linked to the purity of the RIS $\rho_S$: $D_{\mathrm{HS}}(\rho_S,\rho_0) = \trace(\rho_S^2) - 1/(N+1)$. Hence, up to a constant the curves of Fig.~\ref{fig:purity} also represent the evolution of the mean purity in the RIS w.r.t.\ the ancilla parameters.

\begin{figure*}
    \centering
    \includegraphics[width=1\linewidth]{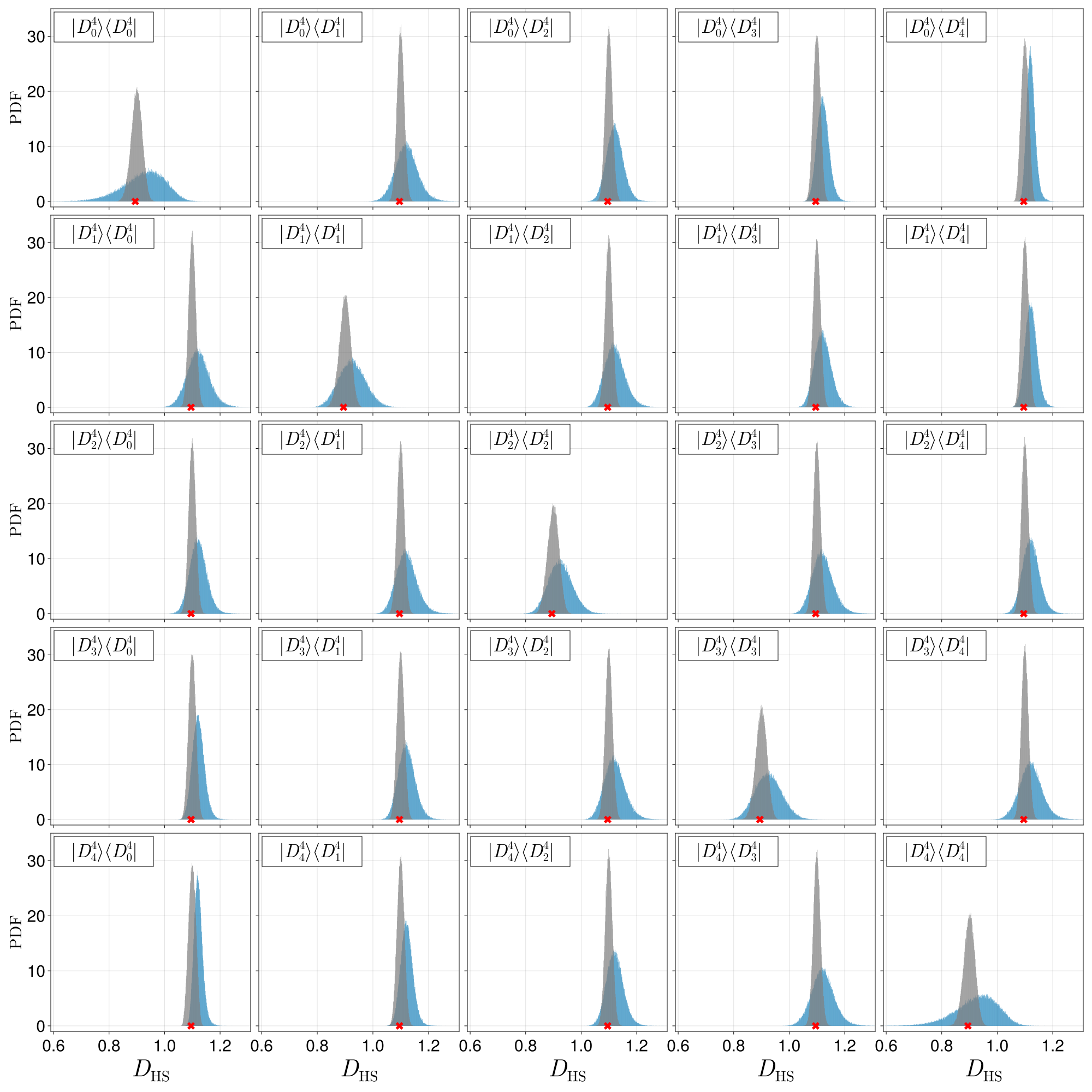}
    \caption{\textbf{Spreading of bound entangled RIS.} Probability density functions (PDFs) of the Hilbert-Schmidt (HS) distance  $D_{\mathrm{HS}}$ between a bound entangled RIS and the Dicke projectors $\ket{D^{\alpha}_N} \langle D^{\beta}_N| $ ($\alpha,\beta= 0,\dotsc,N$) for $N=4$, $N_a = 12$ (MI, blue) and $d_a = 80$ (MII, grey). In each graph, the red cross indicates the HS distance between the maximally mixed state $\rho_{0}$ and the corresponding Dicke projector.
    }
    \label{fig:tomo}
\end{figure*}

\begin{figure*}
    \centering
    \includegraphics[width=1\linewidth]{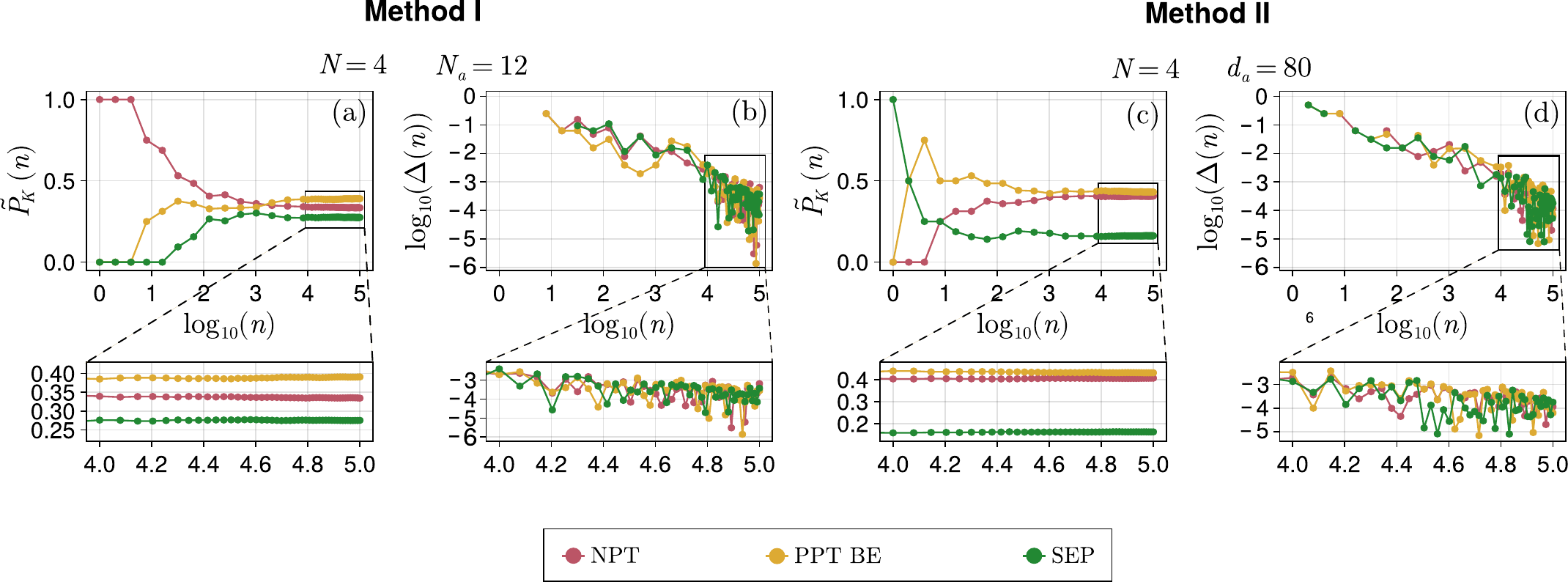}
    \caption{\textbf{Empirical probabilities.} (a) and (c): Empirical probabilities $\tilde{P}_K(n)$ ($K$ = NPT, PPT BE, and SEP) with respect to the number $n$ of RIS generated for $N=4$ for Method I (a) and II (c). The ancilla parameters are $N_a = 12$ for Method I and $d_a = 80$ for Method II. The plot points represent the empirical probabilities computed for the sequence of sample numbers $n = 1, 2, 4, 8, \ldots, 8192$ and then from $n = 10^4$ to $10^5$ by constant steps of 2000 (continuous lines are only guides for the eyes). (b) and (d): Absolute values of the successive differences of the plot points : $\Delta(n_i) = \tilde{P}_K(n_i) - \tilde{P}_K(n_{i-1})$, with $i$ indexing the successive plot points for Method I (b) and II (d). The lower 4 panels display a zoom in the region $n\in [10^4,10^5]$.}
    \label{fig:CONV}
\end{figure*}

We then investigated the variety of the created random induced bound entangled states. To that end, we created samples of random induced PPT BE states of $N = 4,5$ and $6$ qubits using both methods. The ancilla parameters were fixed around the ones giving the maximum probability to get a bound entangled state. For Method I, the selected parameters were $N_a = 12,18$, and $24$ for $N = 4,5$ and $6$ respectively, and $d_a = 100$, $250$, and $300$ for Method II. Figure~\ref{fig:DistBE} displays the Probability Density Functions (PDFs) of the HS distance between two PPT BE RIS for these selected parameters. The PDFs have been estimated from a sample of 30~000 bound entangled RIS that generated a sample of $(30\ 000^2 -30\ 000)/2 = 449\ 985\ 000$ relative distances. In Fig.~\ref{fig:tomo}, we show the PDFs of the HS distance between PPT BE RIS of $N=4$ qubits and the Dicke projectors $\ket{D^{\alpha}_N} \langle D^{\beta}_N| $ ($\alpha,\beta= 0,\dotsc,N$). These latter PDFs have been estimated from a sample of $10^5$ bound entangled RIS.  The figure shows how the two methods generate bound entangled RIS distributed around the maximally mixed state. The distribution is narrower around the MMS for Method II than for Method~I.

\subsection{Comparative analysis of Method I and II}

The results presented in the previous section shed light on the distinct features of the two methods for generating PPT BE states. First, for a fixed number of qubits $N$, and in particular for $N = 4, 5,$ and $6$, Method I attains a slightly lower maximal probability of producing a PPT BE states than Method II. However, as $N$ increases, this difference becomes negligible, and both methods converge toward nearly identical maximal probabilities close to unity.

A striking difference arises in the range of the control parameter over which PPT BE states are observed with high probability. MII possesses a considerably wider parameter region in which the probability of finding PPT BE states exceeds that of NPT or SEP ones. This broad stability means that MII is less sensitive to parameter variations, offering a more robust performance that could facilitate experimental implementation. The ability to maintain a high generation probability over an extended parameter range may prove advantageous in practical situations where fine control of the ancilla parameter is difficult to achieve.

Conversely, MI exhibits a pronounced advantage in terms of the diversity of the generated states. The ensemble of PPT BE states obtained from MI is significantly more varied as shown in Figs.~\ref{fig:DistBE} and~\ref{fig:tomo}. This enhanced diversity may be important for theoretical studies aiming to characterize the structure, geometry, or classification of bound entanglement, or for applications requiring an heterogeneous sample of entangled states.

Taken together, these observations suggest that the choice between MI and MII depends on the intended application. If one prioritizes variety and seeks to probe the richness of the PPT BE landscape, MI constitutes the preferred option. If instead the goal is to achieve higher generation probabilities and to benefit from greater robustness to parameter fluctuations -- features especially relevant in experimental contexts -- MII should be favored. For larger system sizes, where both methods reach nearly maximal probabilities, this distinction becomes less critical; however, in the low-$N$ regime, it remains a significant consideration.

\section{Conclusion} \label{sec:Conclusion}

In this work, we have shown that symmetric random induced states (RIS) offer a practical and efficient route to generating bound entangled states, even in low-dimensional systems, as long as the number of qubits satisfies $N>3$. By considering two distinct methods for generating RIS, namely partial tracing of symmetric multiqubit pure states (method MI) and tracing out a qudit ancilla (method MII), we demonstrated that bound entanglement emerges naturally for suitable choices of the ancilla parameters. The probabilities of generating such PPT bound entangled states increase rapidly with the system size, reaching values close to unity, thus providing a highly efficient mechanism for producing large ensembles of bound entangled states without complex optimization procedures.

The results presented here highlight notable patterns in the distribution of PPT bound entanglement across bipartitions, including a sawtooth behavior in the maximal PPT BE$
_{\textrm{all}}$ probability for odd versus even numbers of qubits and the sequential appearance of PPT entanglement across increasing subsets of bipartitions. Such insights deepen our understanding of multipartite symmetric entanglement and its intricate structure. Furthermore, our exploration of the Hilbert-Schmidt distances provides a quantitative characterization of the diversity of the generated states, illustrating the versatility of symmetric RIS as a resource for entanglement studies.

Our comparative analysis of the two methods reveals a complementary picture. Method I produces a broader variety of bound entangled states, which is valuable for theoretical studies aimed at exploring the structure and geometry of entanglement. Conversely, Method II yields states that are more concentrated around the maximally mixed state, providing higher robustness and a wider parameter range where bound entanglement occurs, which may be advantageous for experimental implementations. These distinctions underscore the importance of selecting a method according to the intended application, whether it be for fundamental research, entanglement characterization, or experimental realizations.

Looking forward, our study opens multiple avenues for future research. One promising line of inquiry is to explore the operational utility of the RIS in quantum information processing, metrology, and quantum communication protocols using both methods. Additionally, extending this approach to higher-dimensional systems and other symmetry classes could further enrich the toolkit for the generation and characterization of bound entanglement.

Overall, our work establishes symmetric RIS as a flexible and powerful framework for the generation of bound entangled states, bridging theoretical analysis and potential experimental realization. The combination of high probability generation, controllable state properties, and methodological simplicity positions this approach as a valuable resource for the ongoing exploration of the subtle phenomenon of bound entanglement in quantum systems.

\paragraph*{Acknowledgments} JL thanks Eduardo Serrano-Ensástiga, Jérôme Denis, and Thomas Michel for fruitful discussions. TB acknowledges
financial support through IISN convention 4.4512.08. Computational resources were provided by the Consortium des Equipements de Calcul Intensif (CECI), funded by the Fonds de la Recherche Scientifique de Belgique (F.R.S.-FNRS) under Grant No. 2.5020.11.

\appendix
\section{Empirical probabilities} \label{app:conv}

Figure \ref{fig:CONV} displays the empirical probabilities $\tilde{P}_K(n)$ ($K$ = NPT, PPT BE, and SEP) with respect to the number $n$ of RIS generated for $N=4$, $N_a = 12$ (Method I) and $d_a = 80$ (Method II). 
At $n = 10^5$, the empirical probabilities are expected to yield an approximation of the exact probabilities to within a precision of $10^{-3}$.

\bibliography{bibliography.bib} 
	
\end{document}